\documentclass[a4paper,12pt]{article}

\usepackage{graphicx}
\usepackage[english]{babel}
\usepackage{amsmath}
\usepackage{amsfonts,amssymb}
\usepackage{algorithm}
\usepackage{algorithmic}

\newcommand{\bZ}{{\bf Z}}
\newcommand{\bJ}{{\bf J}}
\newcommand{\bG}{{\bf G}}
\newcommand{\bU}{{\bf U}}
\newcommand{\bV}{{\bf V}}
\newcommand{\bX}{{\bf X}}
\newcommand{\bA}{{\bf A}}
\newcommand{\bC}{{\bf C}}
\newcommand{\bT}{{\bf T}}
\newcommand{\bE}{{\bf E}}
\newcommand{\bS}{{\bf S}}
\newcommand{\bD}{{\bf D}}
\newcommand{\bQ}{{\bf Q}}
\newcommand{\bI}{{\bf I}}

\newcommand{\bz}{{\bf z}}
\newcommand{\bx}{{\bf x}}
\newcommand{\ba}{{\bf a}}

\newcommand{\tbX}{\widetilde {\bf X}}
\newcommand{\tbA}{\widetilde {\bf A}}
\newcommand{\tbC}{\widetilde {\bf C}}
\newcommand{\tda}{{\tilde a}}
\newcommand{\tdc}{{\tilde c}}

\newcommand{\tbx}{\tilde {\bf x}}

\newcommand{\egaldef}{\stackrel{\mathrm{def}}{=}}

\title{ Orthogonal rotation in PCAMIX\footnote{Submitted paper, August 2011} }

\author{Marie Chavent$^{1,2}$\footnote{marie.chavent@u-bordeaux2.fr}, Vanessa Kuentz$^{3}$ and J\'er\^ome Saracco$^{2,4}$}

\date{}

\begin{document}

\maketitle

\baselineskip=20pt

\begin{center}
\small $^1$ Universit\'e de Bordeaux, IMB, CNRS, UMR 5251, France\\
 \smallskip
 \small $^2$ INRIA Bordeaux Sud-Ouest, CQFD team, France\\
  \smallskip
 $^3$ CEMAGREF, UR ADBX, France\\
 \smallskip
  \small $^4$ Institut Polytechnique de Bordeaux, France\\
\end{center}

\begin{abstract}
Kiers (1991) considered the orthogonal rotation in PCAMIX, a principal component method for a mixture of qualitative and quantitative variables. PCAMIX  includes the ordinary principal component analysis (PCA) and multiple correspondence analysis (MCA) as special cases.  In this paper, we give a new presentation of PCAMIX where the principal components and the squared loadings are obtained from a Singular Value Decomposition.  The loadings of the quantitative variables and the principal coordinates of the categories of the qualitative variables are also obtained directly. In this context, we propose a computationaly efficient procedure for varimax rotation in PCAMIX and a direct solution for the optimal angle of rotation.  
A simulation study shows the good computational behavior of the proposed algorithm. An application on a real data set illustrates the interest of using rotation in MCA. All source codes are available  in the R package ``PCAmixdata''.

\end{abstract}

\section{Introduction}
Kaiser (1958) introduced the  varimax criterion  for the attainment of simple structures by orthogonal rotation in Principal Component Analysis (PCA) .
 This criterion aims at maximizing the sum over the columns of the squared elements of the loading matrix. The loading matrix plays  a significant part in the interpretation of the results since it contains the correlations between the variables and the principal components.  The idea is to get components so that the interpretation is easier, that is to rotate the loading matrix and the standardized principal components so that the groups of variables appear: having high loadings on the same component, moderate ones on a few components and negligible ones on the remaining components.  Because the Singular Value Decomposition (SVD) approach in PCA gives one the freedom for orthogonal rotation, the percentage of variance explained is redistributed along the newly rotated axes, while still conserving the variance explained by the solution as a whole.  

Kiers (1991) extended the varimax criterion for the attainment of simple structures in PCAMIX, a principal component method for the mixture of qualitative and quantitative variables. For qualitative variables, the coefficient used to express the link between a variable and a component is the correlation ratio;  this correlation ratio plays the role of a squared loading. The varimax criterion  is then expressed  with squared loadings defined as correlation ratios for qualitative variables and squared correlations for quantitative variables. Algorithms devised for the determination of an optimal orthogonal rotation in the context of PCA, as proposed for example by Kaiser's (1958), Neudecker (1981) or Jennrich (2001) did not apply to this extended varimax criterion. So Kiers (1991)  proposes a matrix reformulation of this new varimax criterion in order to replace the optimization problem with a problem of simultaneous diagonalization of a set of symmetric matrices  (ten Berge, 1984), and  suggests the use of the algorithm of de Leeuw and Pruzansky (1978) to solve the latter. To the best of our knowledge, the   resulting algorithm has never been presented in a single paper, so we have recalled for comparison purpose the main steps of the matrix reformulation and the simultaneous diagonalization. We shall refer to this algorithm as Kiers' (1991) original approach  to PCAMIX.

In this paper we will first present a new formulation of PCAMIX. It is similar to that of  Escofier (1979) and Pag\`es (2004) in the way  quantitative and qualitative variables are  transformed,  but it is presented  via  a SVD. This presents a direct way to determine both the component scores and the squared loadings and also the principal coordinates of the categories of the qualitative variables as well as the loadings of the qualitative variables. 
Then we will search for an optimal rotation for the PCAMIX varimax criterion using the iterative procedure  suggested by Kaiser (1958) for PCA: we will rotate pairs of dimensions according to an optimal angle $\theta$, iteratively until the process converges. A new direct, specific to PCAMIX determination of this angle is proposed. 
 We shall refer to the resulting algorithm as the SVD approach to PCAMIX.
This algorithm leads to the same final rotation as Kiers' (1991) original approach, however a simulation study shows that it is computationally more efficient.
When all the variables are quantitative, the new algorithm reduces to the classical Kaiser's (1958)  procedure for orthogonal rotation in PCA with a new direct expression of the optimal planar angle $\theta$.

Notice that Kaiser's varimax rotation procedure does not always produce an optimal rotation in PCA. ten Berge (1995) made suggestions for addressing this point for PCA.  This is an open problem for PCAMIX.

This paper is organized as follows. Section \ref{sec2} recalls Kiers' original PCAMIX method and proposes an alternative formulation using SVD. Section \ref{sec3} deals with varimax rotation in PCAMIX. The optimization problem is given section  \ref{optipb}. The determination of the optimal angle of rotation with Kiers' matrix reformulation approach is described section \ref{kier} for purpose of comparison with the direct solution proposed section \ref{nous}. The complete procedure for orthogonal rotation in more than two dimensions is given section \ref{sec:procedure}. 
 A simulation study compares section \ref{tps} the computational time of the proposed rotation procedure with the rotation procedure based on Kiers (1991). In section  \ref{appli} a real data application illustrates the interest of rotation in MCA  and shows some of the outputs and graphical representations available in the R package ``PCAmixdata" we have developed.

\section{The PCAMIX method} \label{sec2}

Let us first introduce some notations used in the presentation of the PCAMIX method.
\begin{itemize}
\item Let $n$ denote the number of observation units, $p_1$ the number of quantitative variables, $p_2$ the number of qualitative variables and $p=p_1+p_2$ the total number of variables. 
\item Let $\bz_j$ be the column vector which contains the standardized scores of the $n$ objects on variable $j$  if the $j$-th variable is quantitative.
\item  Let $\bG_j$ be the indicator matrix for the variable $j$ if the $j$-th variable is qualitative and let  $\bD_j$ be the diagonal matrix of frequencies of categories of this variable. 
\item Let us denote by $m$ the number of categories of the $p_2$ qualitative variables. 
\item Let  $\bG=(\bG_1|\cdots|\bG_j|\cdots|\bG_{p_2})$ be the $n \times m$ matrix of the indicator variables of the $m$ categories of the $p_2$ qualitative variables and let $\bD=\mbox{diag}(\bD_1,\dots,\bD_j,\dots,\bD_{p_2})$ be the $m\times m$ diagonal matrix of frequencies of the $m$ categories.
\item Let $\bJ= \bI_n - \mathbf{1} \mathbf{1}^\prime/n$ be the centering operator where $\bI_n$ denotes the $n\times n$ identity matrix and $\mathbf{1}$ the vector of order $n$ with unit entries.
\end{itemize}

In the two following subsections, we give two formulations of the PCAMIX method and highlight their main differences.

\subsection{The original PCAMIX procedure}

Suppose $k$ is the number of components required in PCAMIX. In Kiers (1991), the procedure computes the $n\times k$ matrix $\bX$ of the standardized component scores, the variance of each component and the $p\times k$ matrix $\bC$ of the squared loadings. The squared loadings are defined as squared correlation for quantitative variables and as correlation ratio for qualitative variables. This procedure is carried out according to the following steps:
\begin{enumerate}
\item  For   {$j=1,\ldots,p$}: calculate the so-called $n \times n$  quantification matrix $\bS_j$ with:
$$
\left\{
\begin{array}{ll}
\bS_j=\frac{1}{n}\bz_j \bz_j^{'} & \mbox{ if variable } j  \mbox{  is quantitative},\\
\bS_j=\bJ \bG_j \bD_j^{-1} \bG_j^\prime \bJ & \mbox{ if variable } j  \mbox{ is qualitative}.
\end{array}
\right.
$$
\item Calculate the $n \times n$ matrix $\bS=\sum_{j=1}^p \bS_j$.
\item Perform an EigenValue Decomposition of $\bS$. The matrix $\bX$ of the standardized component scores  is given by the  first $k$ eigenvectors of $\bS$ normalized to $n$ (such that $\bX'\bX=n\bI_k$).
\item For  $l=1,\ldots,k$: calculate the variance of the $l$-th component given by $\bx_l'\bS\bx_l$ where $\bx_l$ denotes the $l$-th column of $\bX$.
\item Calculate the matrix $\bC$ of the  squared loadings of the $p$ variables on the $k$ components with $c_{jl}=\frac{1}{n}\bx_l' \bS_j \bx_l$.  For quantitative (resp. qualitative) variables, $c_{jl}$  is the squared correlation (resp. correlation ratio) between the variable $j$ and the component $l$.
\end{enumerate}
When all the variables are quantitative (resp. qualitative), this procedure is equivalent to PCA (resp. MCA). But  the loadings  (the correlations between the variables and the components) and  the principal coordinates of the categories  (the barycenters of the component scores) are not directly provided and must be calculated afterwards if desired.  From a practical point of view this procedure requires the construction and the storage of  $p$ matrices of dimension $n \times n$ which can leads to  memory size problems when $n$ and $p$ increase.

\subsection{The SVD based PCAMIX procedure  } \label{svd-pcamix}

This procedure is carried out according to the following steps:

\begin{enumerate}
\item Determine the $n \times (p_1+m)$ matrix of interest $\bZ=\frac{1}{\sqrt{n}}(\bZ_1|\bZ_2)$ where :
\begin{itemize}
\item $\bZ_1=(\bz_1|\cdots|\bz_j|\cdots|\bz_{p_1})$ is the $n \times p_1$ matrix of the standardized scores of the $n$ observation units (objects) on the $p_1$ quantitative variables. 
\item $\bZ_2$ is the  $n \times m$ matrix obtained by recoding  $\bG$ in the following way:  $\bZ_2=\bJ\bG\bD^{-1/2}$.
\end{itemize}
\item Perform the SVD of $\bZ$ :
\begin{equation}\label{svd}
\bZ=\bU \Lambda \bV',
\end{equation} 
 where $\bU'\bU=\bV'\bV=\bI_r$, $\Lambda$ is the diagonal matrix of singular values (in weakly descending order) and $r$ is the rank of $\bZ$. 
\item Calculate the $n\times k$ matrix  of the standardized component scores: \begin{equation}\label{matX}
\bX=\sqrt{n}\bU_k
\end{equation} 
 where $\bU_k$ denotes the matrix of the first  $k$ columns of $\bU$.
\item For   {$\ell=1,\ldots,k$}, the standard deviation of the $\ell$-th component is given by the $\ell$-th singular value in $\Lambda$.
\item  Calculate the matrix: 
\begin{equation}\label{matA}
\bA=\bV_k\Lambda_k,
\end{equation} 
where  $\bV_k$ denote the matrix of the first  $k$ columns of $\bV$ and $\Lambda_k $ the diagonal matrix of the $k$ largest singular values.
\item Write $\bA=\left(\frac{\bA_1}{\bA_2}\right)$ the concatenation of a $p_1 \times k$ matrix $\bA_1$ and a $m\times k$ matrix $\bA_2$. 
\begin{itemize}
 \item The matrix $\bA_1$ contains the loadings of the quantitative variables (the correlations between the quantitative variables and the components).
\item The matrix $\bD\bA_2$ contains the principal coordinates of the categories of the qualitative variables.
\item  Calculate the matrix $\bC$ of the  squared loadings of the $p$ variables on the $k$ components. This matrix is obtained from the matrix $\bA$ as follows:
$$
\left\{
\begin{array}{ll}
c_{jl}=a_{jl}^2 & \mbox{ if variable } j  \mbox{  is quantitative},\\
c_{jl}=\sum_{s \in I_j} a_{sl}^2 & \mbox{ if variable } j  \mbox{ is qualitative},
\end{array}
\right.
$$
where $I_j$ is the set of row indices of $\bA$ associated with the categories of the qualitative variable $j$. To simplify the notations,  we note hereafter $c_{jl}=\sum_{s \in I_j} a_{sl}^2$ for both quantitative and qualitative variables with $I_j=\{j\}$ in the quantitative case.
\end{itemize}
\end{enumerate}

Note that the matrix $\bX$ of the standardized component scores is obtained  from the SVD of the recoded data matrix $\bZ$ whereas it was obtained from the Eigenvalue Decomposition of the matrix $\bS$ (the sum of the quantification matrices $\bS_j$) in Kiers' original approach. Also, the matrix $\bC$ of the squared loadings (squared correlations or correlation ratios between the variables and the components) is calculated here from the only matrix $\bA$ obtained with the SVD of  $\bZ$ whereas it was calculated from the  two matrices $\bX$ and $\bS_j$ in Kiers' original approach.

Contrary to the original PCAMIX approach, this procedure simultaneously provides  the loadings of the quantitative variables and  the principal coordinates of the categories of the qualitative variables.  Moreover, when the data are mixed (quantitative and qualitative),  the well known barycentric property  in MCA remains true: the coordinates of the categories are the averages of the standardized component scores of the objects in those categories. The matrices $\bX$, $\bA_1$ and $\bD\bA_2$  are then used to plot the observation units, the quantitative variables and the categories  with the same interpretation rules as in PCA and MCA. Matrix $\bC$ is  used to plot the quantitative and  qualitative variables on a same graphic.

\section{Varimax rotation in PCAMIX} \label{sec3}

\subsection{The optimization problem} \label{optipb}

\paragraph{Why using rotation ?} As shown by Eckart and Young (1936), from the SVD in (\ref{svd}) and definitions of matrices $\bX$ and $\bA$ given in (\ref{matX}) and (\ref{matA}), the matrix $\bX\bA'$ is a rank $k$ least squares approximation of $\bZ$.
Let us introduce $\bT$ an orthonormal rotation matrix: $\bT\bT'=\bT'\bT=\bI_k$.
Let  $\tbX=\bX\bT$ and $\tbA=\bA\bT$.
As $\bX\bA'=\tbX\tbA'$,  this approximation is not unique over orthogonal rotations. 

This non-uniqueness can be exploited to improve the interpretability of the original solutions. To simplify the interpretations, the matrices $\bX$ and  $\bA$ are then  rotated in such a way that when considering one variable, few squared loadings are large (close to 1) and as many as possible are close to zero.

\paragraph{The varimax problem.}
In PCA, since $\tbA$ contains the loadings of the variables after rotation, the varimax rotation problem is formulated as 
\begin{equation}
\begin{array}{ll}
\displaystyle \max_{\bT} & f(\bT), \\
\mbox{s.t. }& \bT \bT'=\bT' \bT=\bI_k,
\end{array}
 \label{pboptim}
\end{equation}
where
\begin{equation}
f(\bT)=\sum_{l=1}^k\sum_{j=1}^p (\tda_{jl}^2)^2 - \frac{1}{p} \sum_{l=1}^k  \left(\sum_{j=1}^p \tda_{jl}^2 \right)^2
\label{varipca}
\end{equation}
is the varimax function measuring the simplicity of the components after rotation.

In the SVD approach of PCAMIX, the varimax function $f$ is defined   by replacing in  (\ref{varipca}) the  terms $\tda_{jl}^2$ by  $\tdc_{jl}$, where the $\tdc_{jl}=\sum_{s \in I_j} \tda_{sl}^2$ are  the squared loadings after rotation:
\begin{equation}
f(\bT) =\sum_{l=1}^k\sum_{j=1}^p (\tdc_{jl})^2 - \frac{1}{p} \sum_{l=1}^k  \left(\sum_{j=1}^p \tdc_{jl}\right)^2. \label{varipcamix} 
\end{equation}
Note that the squared loadings  after rotation $\tdc_{jl}$ are squared correlations (resp. correlation ratios)  between the quantitative (resp. qualitative) variables and  the rotated components.

\medskip

For comparison purpose, we recall Kiers' original expression of the varimax function in PCAMIX:  the squared loadings after rotation $\tdc_{jl}$ are given by $\frac{1}{n}\tbx_l' \bS_j \tbx_l$,  where $\tbx_l$ denotes the $l$-th column of $\tbX$. Hence the varimax function (\ref{varipcamix}) becomes:
\begin{equation}
f(\bT)=\sum_{l=1}^k\sum_{j=1}^p \left(\frac{1}{n}\tbx_l' \bS_j \tbx_l\right)^2 - \frac{1}{p} \sum_{l=1}^k  \left(\sum_{j=1}^p \frac{1}{n}\tbx_l' \bS_j \tbx_l\right)^2.
\label{varikiers}
\end{equation}

\paragraph{The iterative optimization procedure.}
Because a direct solution for the optimal $\bT$ is not available, an iterative optimization procedure 
 suggested by Kaiser (1958) for PCA can be used for PCAMIX. 
 The idea is to consider at each iteration a planar rotation for which the rotation matrix $\bT$ only depends of an angle $\theta$ (see below for details).
 This procedure rotates  pairs of dimensions in the following way: the single-plane rotations are applied to dimensions 1 and 2, 1 and 3, $\ldots$, 1 and $k$, 2 and 3,$\ldots$, $(k-1)$ and $k$, iteratively until the process converges, i.e. until ${k(k-1)}/{2}$ successive rotations providing an angle of rotation equal to zero are obtained.

The key point of this rotation procedure is  the definition of the single-plane rotation step. We give next details on the calculation of the optimal angle for planar rotation. Then we give the complete iterative procedure for rotation in more than two dimensions.

\subsection{Planar rotation}
 \noindent
 Single planar rotations are obtained with a rotation matrix $\bT$ defined by
\begin{equation}
\bT=
\left[
\begin{array}{cc}
\text{cos }\theta & -\text{sin }\theta \\
\text{sin }\theta & \text{cos } \theta
\end{array}
\right]
\label{matTdim2}
\end{equation} 
where $\theta$ is the angle of rotation.
The varimax rotation problem \eqref{pboptim} is then rewritten as:
$$
\displaystyle \max_{\theta \in \mathbb{R}}f(\theta).
$$

For purpose of comparison we recall first the solution based on Kiers' matrix reformulation before we give our direct solution. 

\subsubsection{Planar rotation using the Kiers' matrix reformulation} \label{kier}
Kiers (1991) proposes to use a procedure of simultaneous diagonalization of a set of symmetric matrices (ten Berge, 1984; de Leeuw and Pruzansky,1978)  to solve the global  varimax optimization problem  (\ref{pboptim}). For that purpose he gives the following matrix reformulation of the formula~(\ref{varikiers}) giving $f$ :
\begin{equation}
f(\bT)=p^{-2}\sum_{j=1}^{p}\mbox{Trace}\left(\bT^\prime \bE_j\bT(\mbox{Diag }\;\bT^\prime \bE_j \bT)\right)
\label{varikiers2}
\end{equation}
where
\begin{equation}
\bE_j=p\; \bX'\bS_j\bX-n\Gamma \label{eqE}
\end{equation}
and $\Gamma$ is the diagonal matrix with the $k$ first eigenvalues of   $\bS$ on its diagonal. 

Careful reading of ten Berge (1984) and  de Leeuw and Pruzansky (1978) shows that the procedure for simultaneous diagonalization of the matrices $\bE_j$ is equivalent to Kaiser's iterative optimization procedure  with  the optimal angle $ \theta$ of  single plane rotations defined by the equation:
\begin{equation}
\tan(4 \theta) = \frac{a}{b},
\label{condkiers}
\end{equation}
where 
\begin{equation}\label{akiers}
a=4\sum_{j=1}^p e_{12}^j(e_{11}^j-e_{22}^j) ~~~\mbox{and}~~~
b=\sum_{j=1}^p(e_{11}^j-e_{22}^j)^2-4\sum_{j=1}^p(e_{12}^j)^2 
\end{equation}
and
$\bE_j=\left(\begin{array}{cc} 
 e_{11}^j & e_{12}^j \\
 e_{21}^j &e_{22}^j
\end{array}
\right)
$
is defined in (\ref{eqE}).

As mentionned by several authors (see for instance Nevels, 1986;  ten Berge, 1984; de Leeuw and Pruzansky, 1978 and  Kaiser, 1958)  equation (\ref{condkiers})  is only  a necessary condition obtained upon setting the first order derivative of the objective function to zero. Both Kaiser (1958) and de Leeuw and Pruzansky (1978)  developed a procedure for determining the optimal $\theta$ from the sign of the second order derivative of the objective function. These two procedures, expressed in tabular form, give the appropriate solution  for every possible combination of signs of $a$ and $b$. 

\subsubsection{Planar rotation using the SVD approach of PCAMIX}\label{nous}

The varimax function $f(\bT)$ defined with the SVD approach in (\ref{varipcamix}) is written:
\begin{equation}\label{ftheta}
f(\theta)=   \sum_{j=1}^p \left( \sum_{s \in I_j}  \tda_{s1}^2  \right)^2 + \sum_{j=1}^p \left(\sum_{s \in I_j}  \tda_{s2}^2  \right)^2  -\frac{1}{p} \left(\sum_{j=1}^p \sum_{s \in I_j}  \tda_{s1}^2  \right)^2 - \frac{1}{p}  \left(\sum_{j=1}^p \sum_{s \in I_j}  \tda_{s2}^2 \right) ^2 \end{equation} 
with 
\begin{equation}  \label{as1}
\tda_{s1}=a_{s1}\text{ cos}(\theta)+a_{s2}\text{ sin}(\theta)~~~\mbox{and}~~~ \tda_{s2}=-a_{s1}\text{ sin}(\theta)+a_{s2}\text{ cos}(\theta).
\end{equation} 
This function is equal to (see Appendix):
\begin{equation}
\label{ftheta2}
  f(\theta)=f(0) + \frac{\rho}{4p} \big(\cos(4\theta-\psi) - \cos\psi\big)
\end{equation}
where $\rho$ and $\psi$ are defined by :
\begin{equation}
\label{rhopsi}
  \rho=(a^{2}+b^{2})^{1/2} \quad,\quad \cos \psi=b/\rho \quad,\quad \sin \psi=a/\rho
\end{equation}
with  $a$ and $b$ given by :
\begin{equation}\label{akaiser}
 a= 2p \sum_{j=1}^p {u_j} {v_j} -2\sum_{j=1}^p {u_j} \sum_{j=1}^p {v_j} ~,~~ b= p \sum_{j=1}^p ({u_j}^2-{v_j}^2) -\left( \sum_{j=1}^p {u_j}\right)^2 + \left( \sum_{j=1}^p {v_j}\right)^2,
\end{equation}
where ${u_j}$ and ${v_j}$ are defined by :
\begin{equation}  \label{usvd}
{u_j}= \sum_{s \in I_j} (a_{s1}^2-a_{s2}^2) ~~~\mbox{and}~~~ {v_j}=2\sum_{s \in I_j} a_{s1}a_{s2}\ .
\end{equation}

The function $f$ obtained in (\ref{ftheta2}) is maximum for $\cos(4\theta-\varPsi)=1 \Leftrightarrow  4\theta-\varPsi=2k\pi$, thus the optimal angles are :
\begin{equation}
 \theta=\frac{\varPsi}{4}+k\frac{\pi}{2}, \;\;  k \in \mathbb{Z}. \label{directsol}
\end{equation}

Note that the above expressions of ${u_j}$ and ${v_j}$   contain as special cases (take $I_j=\{j\}$)  those  defined by Kaiser (1958) for the PCA varimax solution. 
Note also that the classical necessary condition ({\ref{condkiers})  immediately follows  by setting  the expression (\ref{dftheta2}) of $pf^{\prime}(\theta)$ given in the Appendix to zero (the coefficients $b$ and $a$ given by (\ref{akiers}) on one side, and (\ref{akaiser})(\ref{usvd}) on the other side are proportional). 

\subsection{The iterative rotation procedure.} \label{sec:procedure}
We consider now the case where the number $k$ of dimensions in the rotation is greater than two. The  iterative rotation procedure gives the matrix $\tbX$ of the rotated standardized component scores and the matrix  $\tbA$ which is used to obtain the rotated squared loadings, the rotated loadings (correlations) of the quantitative variables and the rotated principal coordinates of the categories. This procedure is carried out according to the following steps:

\begin{enumerate}
\item Initialization : $\tbX=\bX$ and $\tbA=\bA$ where the $n \times k$ matrix $\bX$ and the $(p_1+m) \times k$ matrix $\bA$ are given by the SVD based PCAMIX procedure given section  \ref{svd-pcamix} .
\item  For $l=1,\ldots,k-1$ and $t=(l+1),\ldots,k$, calculate for the pair of  dimensions $(l,t)$:
\begin{itemize}
\item[-]  the angle of rotation $ \theta={\varPsi}/{4} $ with $\varPsi$ defined in (\ref{rhopsi}) . We choose:
\begin{equation}
\displaystyle
\varPsi=   \left \{
   \begin{array}{ rll}
   \displaystyle   \text{arcos}(\frac{b}{\sqrt{a^2+b^2}} ) & \text{if} & a \geq 0,\\
   \displaystyle   -\text{arcos}(\frac{b}{\sqrt{a^2+b^2}})  & \text{if} & a \leq 0.
   \end{array}
   \right .
\label{psi}
\end{equation}
where   $a$ and $b$ are defined in (\ref{akaiser}).
\item[-] the matrix of rotation $
\bT=
\left[
\begin{array}{cc}
\text{cos } \theta & -\text{sin } \theta \\
\text{sin } \theta & \text{cos } \theta
\end{array}
\right]
$,
\item[-] the matrices $\tbX$ and $\tbA$ updated by rotation of their $l$-th and $t$-th column. 
\end{itemize}
\item Repeat the previous step until the $k(k-1)/2$ angles $ \theta$ are equal to zero. 
\item Calculate:
\begin{itemize}
\item[-]  the matrix $\tbC$   with  $\tdc_{jl}=\sum_{s \in I_j} \tda_{sl}^2 $.
\item[-]  the matrix $\tbA_1$ of the $p_1$ first rows of $\tbA$ which contains the rotated  loadings of the quantitative variables.
\item[-]  the matrix $\tbA_2$ of the $m$ last rows of $\tbA$ and the matrix  $\bD \tbA_2$ which contains the rotated principal coordinates of the categories of the qualitative variables. 
\end{itemize}
\end{enumerate}

The main differences between this procedure and that constructed with Kiers' matrix reformulation are the following:
\begin{itemize}
\item  The expressions of $a$ and $b$ in step (2):  in this procedure they are expressed according to the matrix $\bA$ of dimension  $(p_1+m) \times n$ where $p_1$ is the number of quantitative variables and $m$ is the total number of categories. With Kiers' matrix reformulation, $a$ and $b$ are expressed according to the $p$ matrices $\bS_j$ of dimension $n \times n$. Then the calculation and the storage of these matrices may be time and space consuming.
\item The direct determination of the optimal angle in step (2). Having  an explicit expression for the solution is of theoretical interest and is more straightforward from a computational point of view. 
\item  The outputs: this procedure provides directly the rotated loadings of the quantitative variables and  the rotated principal coordinates of the categories which are used for graphical representations after rotation.
\end{itemize}

\section{Numerical studies}
The procedure proposed in this paper for varimax orthogonal rotation in PCAMIX  has been implemented in R. A package called ``PCAmixdata'' is already available on the CRAN website. In this section, this algorithm is compared on simulated  data with Kiers' rotation procedure.  Then an application on a real data example illustrates the possible benefits of using rotation in MCA as particular case of PCAMIX. 

\subsection{A simulation study: comparison of computational times} \label{tps}

An iterative rotation procedure based on Kiers' matrix reformulation has also been implemented in R. This procedure is  that proposed section \ref{sec:procedure} with the following modifications:
\begin{itemize}
\item Kiers' original PCAMIX procedure is used in the initialization step in place of the SVD based PCAMIX procedure.
\item  All the calculations and outputs based on the matrix $\bA$ are removed because this matrix is not part of the original PCAMIX procedure.
\item The coefficients $a$ and $b$  in step 2 are calculated according to their expressions (\ref{akiers}) associated to Kiers' matrix reformulation. Note that the ratio $\frac{a}{b}$ is the same with the two approaches (SVD and matrix reformulation) so the optimal angle $ \theta$ is the same.
\item In step 4 the squared loadings are calculated with their expression in the original PCAMIX approach.
\end{itemize}
The computation  time of the two rotation procedures (the one based on Kier's matrix reformulation and the one based on the SVD approach of PCAMIX) is compared from simulated datasets with varying parameters: the number $p$ of variables ($p/2$ quantitative and $p/2$ qualitative) and the number $n$ of observations. For each set of parameters ($n$,\,$p$), 20 simulations are drawn. More precisely the datasets are built using the following procedure:
\begin{itemize}
\item A dataset with $n$ observations and $p$ variables is drawn from a multivariate normal distribution with a covariance matrix $\Sigma=\bQ'\bQ$ where $\bQ$ is a $p\times p$ matrix drawn from a uniform distribution on the interval $[0.2;0.4]$.
\item The $p/2$ last variables are distributed in three equal-count categories. Each dataset is then constituted of $p_1=p/2$ quantitative variable,  $p_2=p/2$ qualitative variable and the total number of categories is $m=3*p/2$.
\end{itemize}
Because the two rotation procedures iterate  planar rotations until convergence, we compare their computation time for $k=2$.  The median computation times  (over the 20 replications)  are given in Table \ref{table:tpscalcul} and the ratio between the computation time of the two approaches are given in Table \ref{table:ratiotpsl}.

\begin{table}[!htb]
\centering
\begin{small}
\begin{tabular}{|c|c|cccc|}
\hline
&&$p$=10&$p$=50&$p$=100&$p$=200\\ \hline
$n$=50 &Matrix reformulation &0.05&0.12&0.22&0.44\\
$n$=50&SVD&0.02&0.06&0.12&0.27 \\ \hline
$n$=100&Matrix reformulation&0.14&0.33&0.56&1.04\\
$n$=100&SVD&0.02&0.09&0.17&0.34 \\ \hline
$n$=200&Matrix reformulation&0.55&1.12&1.86&3.38\\
$n$=200&SVD&0.02&0.11&0.26&0.53\\ \hline
$n$=400&Matrix reformulation&2.15&4.32&7.1&12.65\\
$n$=400&SVD &0.03&0.16&0.37&0.89\\ \hline
$n$=800&Matrix reformulation&10.06&19.27&30.54&error\\
$n$=800&SVD &0.05&0.25&0.58&1.79\\ \hline
\end{tabular}
\end{small}
\caption{Median computation time (in seconds) of two PCAMIX rotation procedures: the one based on Kiers' matrix reformulation and the one based on the SVD appoach.}
\label{table:tpscalcul}
\end{table}

\begin{table}[!htb]
\centering
\begin{small}
\begin{tabular}{|c|cccc|}
\hline
    &   $p$=10 &$p$=50 &$p$=100 &$p$=200\\ \hline
$n$=50   & 2.9&  2.0&   1.8&   1.6 \\ \hline
$n$=100  & 8.7 & 3.8 &  3.3 &  3.0  \\ \hline
$n$=200  &23.2 &10.3 &  7.0 &  6.4  \\ \hline
$n$=400 & 69.4 &27.7  &19.0 & 14.2  \\ \hline
$n$=800 &214.1 &77.4  &52.9 &   error  \\ \hline
\end{tabular}
\end{small}
\caption{Ratio between the median computation time of the two rotation procedures (Matrix reformulation/SVD).}
\label{table:ratiotpsl}
\end{table}

Table \ref{table:tpscalcul} shows that the SVD approach is faster than the matrix reformulation approach for all configurations.  For configurations where $p=10$, Table \ref{table:ratiotpsl} shows that the SVD approach is from 3 times faster  for $n=50$ to   214 times faster for $n=800$. For configurations with greater values of $p$, this ratio is less important but still increases with $n$. For the configuration where $n$ and $p$ are great ($n=800$ and $p=200$) an error occurs with the rotation procedure based on Kiers'matrix refromulation. The maximum capacity of memory size of the computer was reached in that case. This error occurs during the calculation of the $p$ matrices $\bS_j$ of size $n \times n$. This confirms the computational efficiency of the proposed SVD approach.

\subsection{A real data application} \label{appli}
This real data application illustrates the interest of rotation in MCA.  A food habits survey\footnote{This survey was realized by the Bordeaux School of Public Health (Institut de Sant\'e Publique, d'Epid\'emiologie et de D\'eveloppement - ISPED)} was carried out  in 1999 on students living in the region ``Aquitaine'' in south of west France. We focus on the answers of 2885 students to 12 binary questions concerning their consumption at breakfeast (coffe, cereals, eggs...).  The PCAMIX method (equivalent here to MCA)  has been applied to this dataset and the  first 4 components have been rotated.

 In Figure \ref{fig:sload} the association of the variables with the first two components is obviously  easier after rotation. This rotation of the first four components leads in Table  \ref{table:correlationratio} to clear associations between the binary variables: coffe is associated with milk, eggs  with cheese and deli, bread with jam and cereals with pure milk. The effect of  the rotation  on the objects' scores and on the categories' coordinates can also be visualized in Figures \ref{fig:scores} and \ref{fig:categ}. The interpretation rule associated with the barycentric property remains true after rotation.

\begin{table}[!htb]
\centering
\scriptsize
\begin{tabular}{|c|cccc|cccc|}
\hline
&\multicolumn{4}{c|}{\textit{Before rotation}} & \multicolumn{4}{c|}{\textit{After rotation}} \\
   &  1 &  2& 3& 4&  1 &  2& 3& 4\\
   \hline
coffe & 0.23 & 0.22 & 0.06 & 0.05 & {\bf 0.49} & 0.00 & 0.00 & 0.07\\
tea & 0.05 & 0.01 & 0.06 & 0.18 & 0.05 & 0.02 & 0.06 & 0.17\\
milk & 0.15 & 0.16 & 0.08 & 0.00 & {\bf 0.37} & 0.00 & 0.01 & 0.01\\
milk chocolate & 0.43 & 0.18 & 0.01 & 0.06 & {\bf 0.62} & 0.00 & 0.01 & 0.05\\
pure milk & 0.02 & 0.00 & 0.05 & 0.40 & 0.01 & 0.00 & 0.02 & {\bf 0.44}\\
cheese & 0.18 & 0.23 & 0.01 & 0.00 & 0.00 & {\bf 0.42} & 0.00 & 0.00\\
deli & 0.20 & 0.27 & 0.00 & 0.05 & 0.00 & {\bf 0.51} & 0.00 & 0.01\\
eggs & 0.20 & 0.37 & 0.00 & 0.01 & 0.00 & {\bf 0.58} & 0.00 & 0.00\\
jam & 0.06 & 0.02 & 0.49 & 0.02 & 0.00 & 0.00 & {\bf 0.59} & 0.00\\
honey & 0.00 & 0.05 & 0.14 & 0.20 & 0.00 & 0.02 & 0.20 & 0.16\\
bread & 0.11 & 0.01 & 0.45 & 0.00 & 0.01 & 0.01 & {\bf 0.53} & 0.03\\
cereals & 0.01 & 0.01 & 0.12 & 0.22 & 0.05 & 0.01 & 0.04 & {\bf 0.27}\\
\hline
\end{tabular}
\caption{Correlation ratio (squared loadings) between the variables and the  first 4 components before and after rotation}
\label{table:correlationratio}
\end{table}

\begin{figure}[!htb]
\centering
 \includegraphics[scale=0.5]{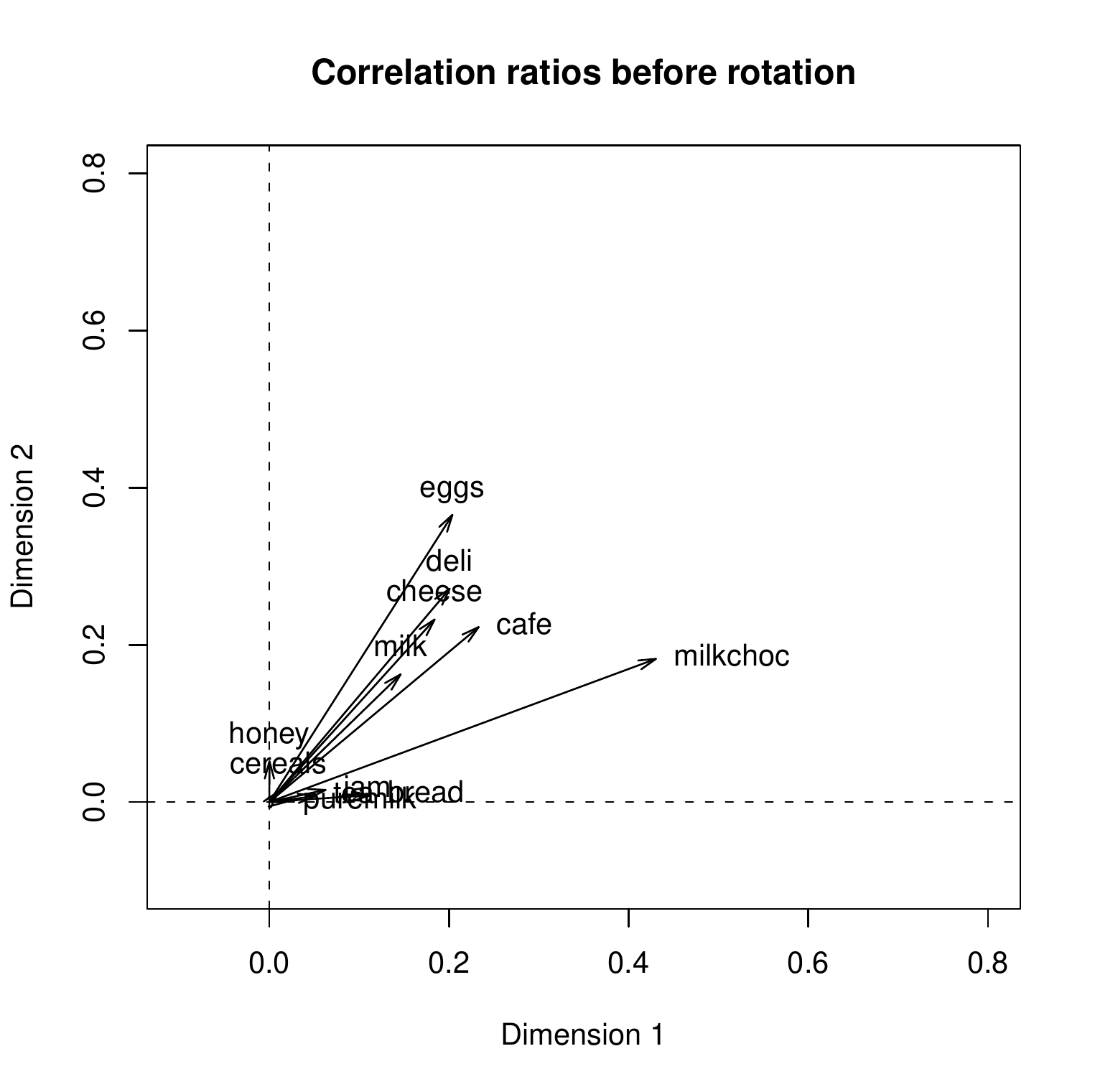}
 \includegraphics[scale=0.5]{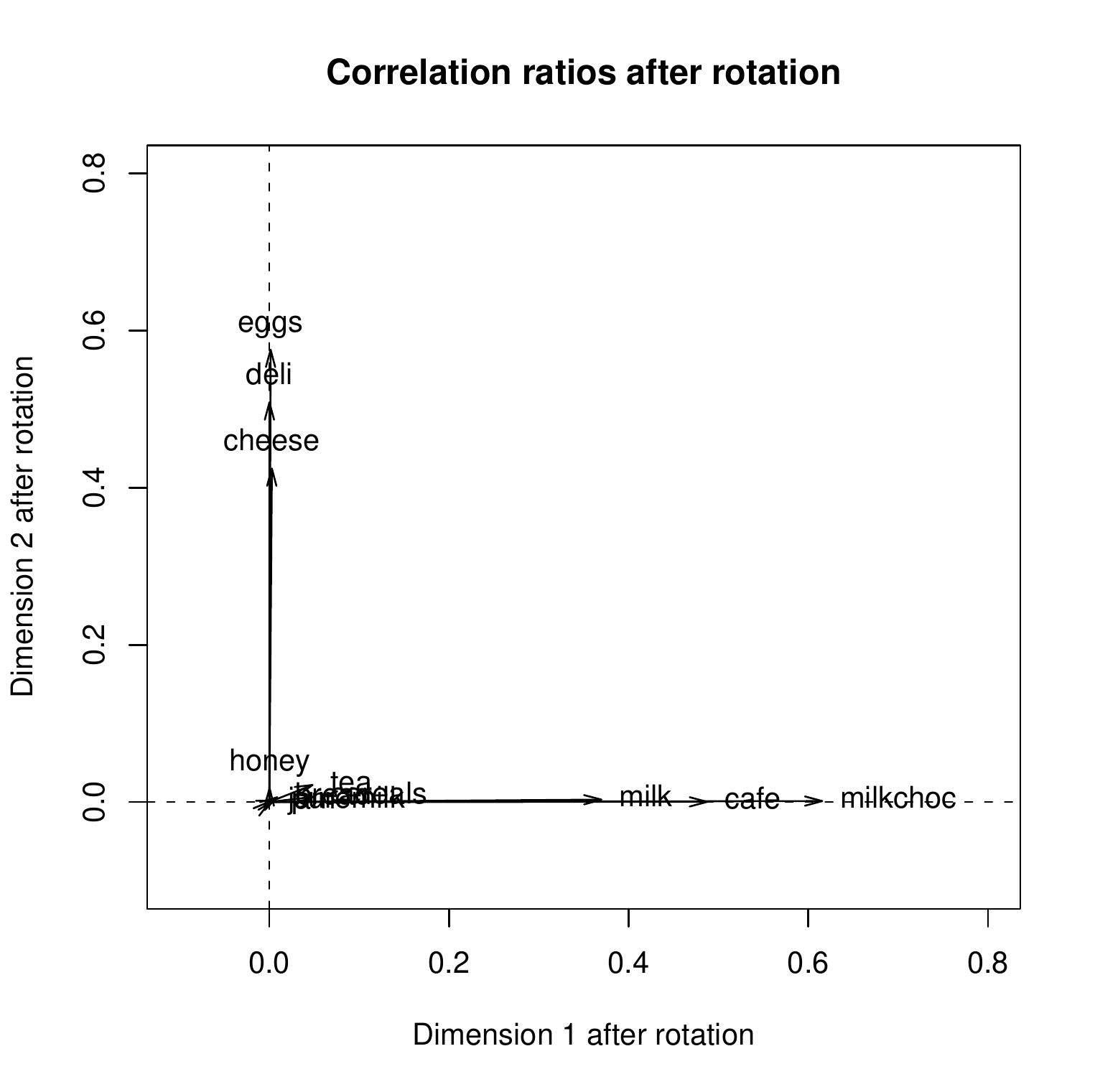}\\
\caption{Plots of the correlation ratios between the variables and the two first components before rotation and after rotation.}
\label{fig:sload}
\end{figure}

\begin{figure}[!htb]
\centering
 \includegraphics[scale=0.5]{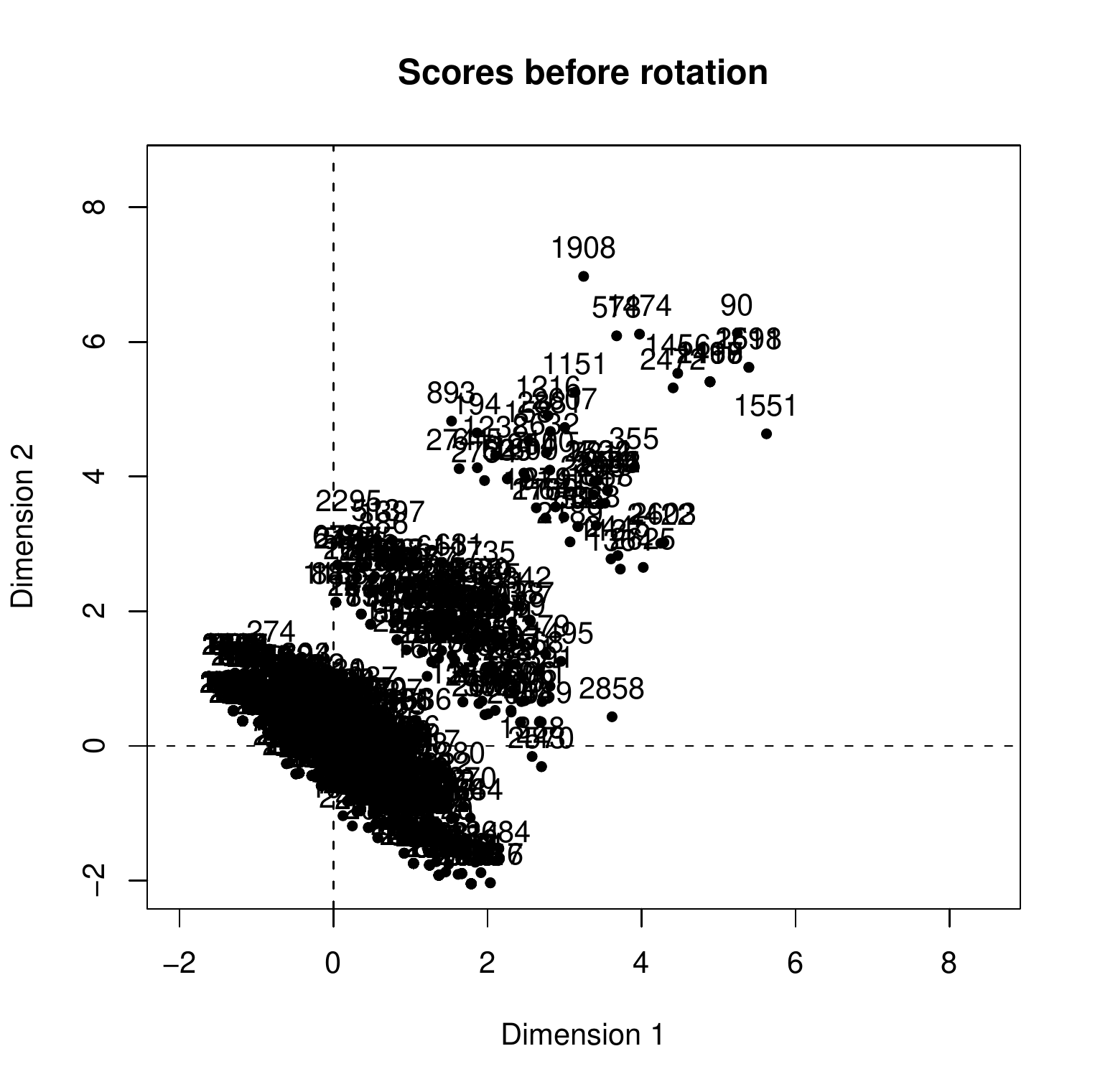}
 \includegraphics[scale=0.5]{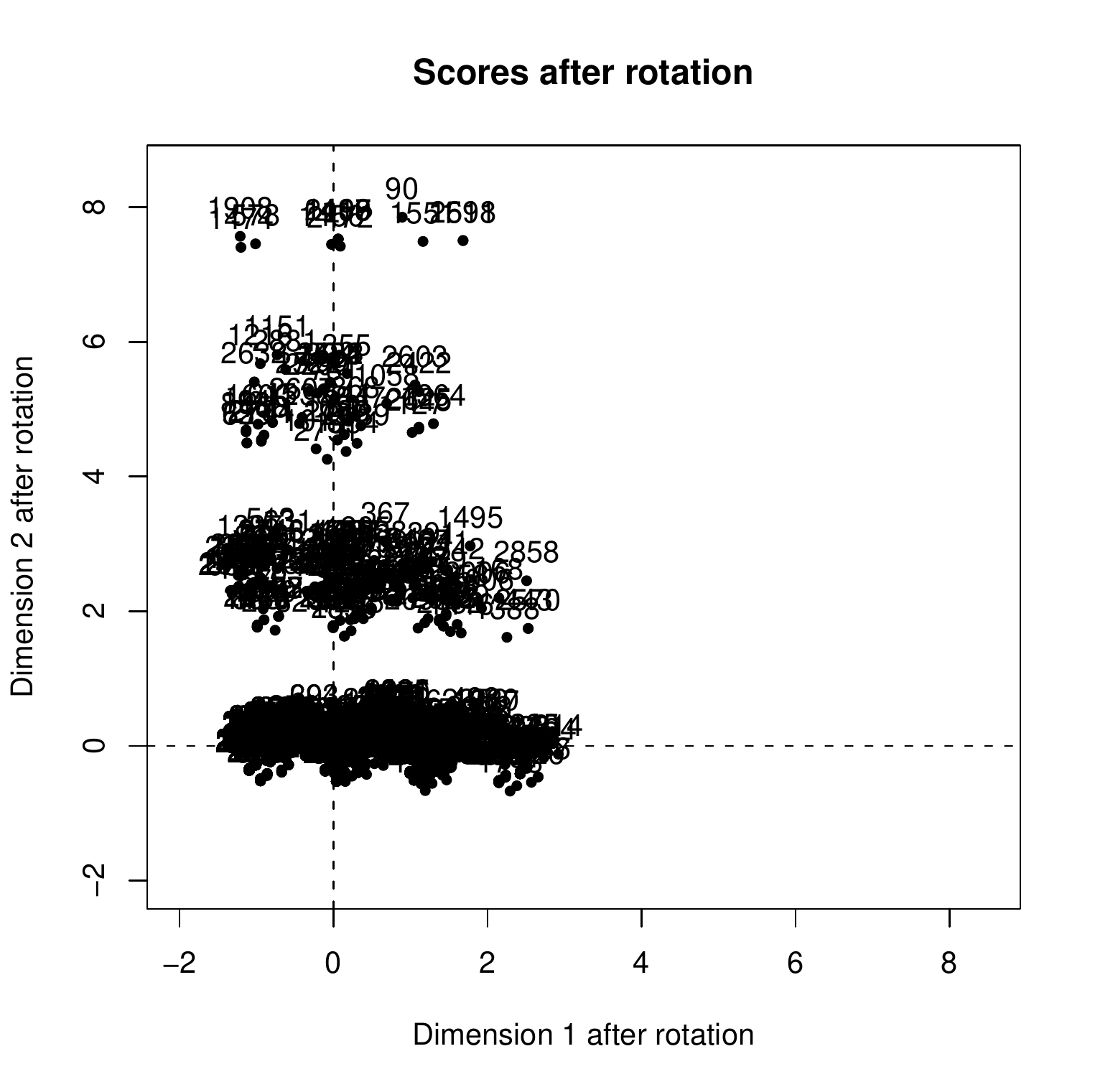}\\
\caption{Plots of the (standardized) scores of the 2885 students on the  first two components before and after rotation. }
\label{fig:scores}
\end{figure}

\begin{figure}[!htb]
\centering
 \includegraphics[scale=0.5]{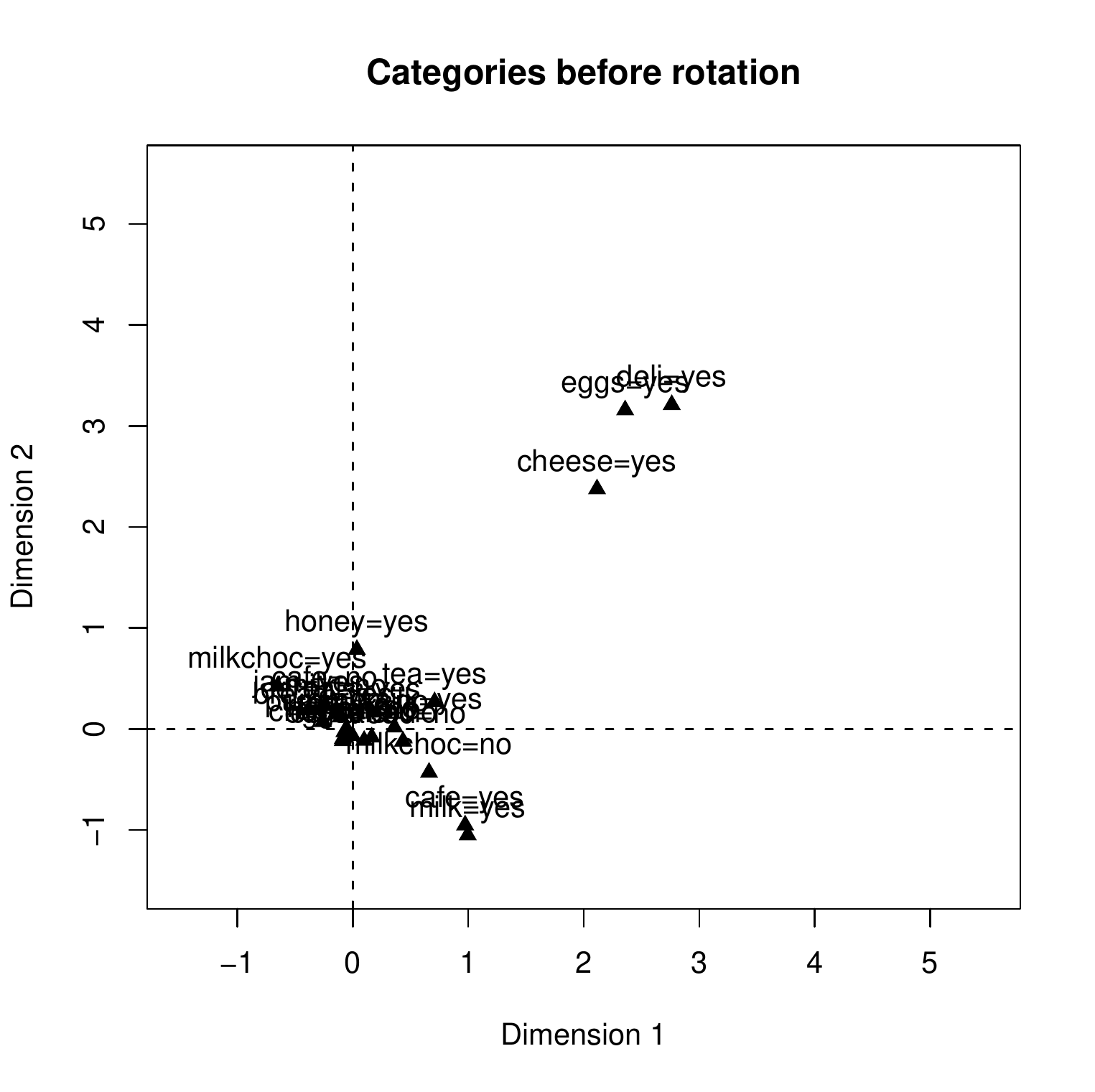}
 \includegraphics[scale=0.5]{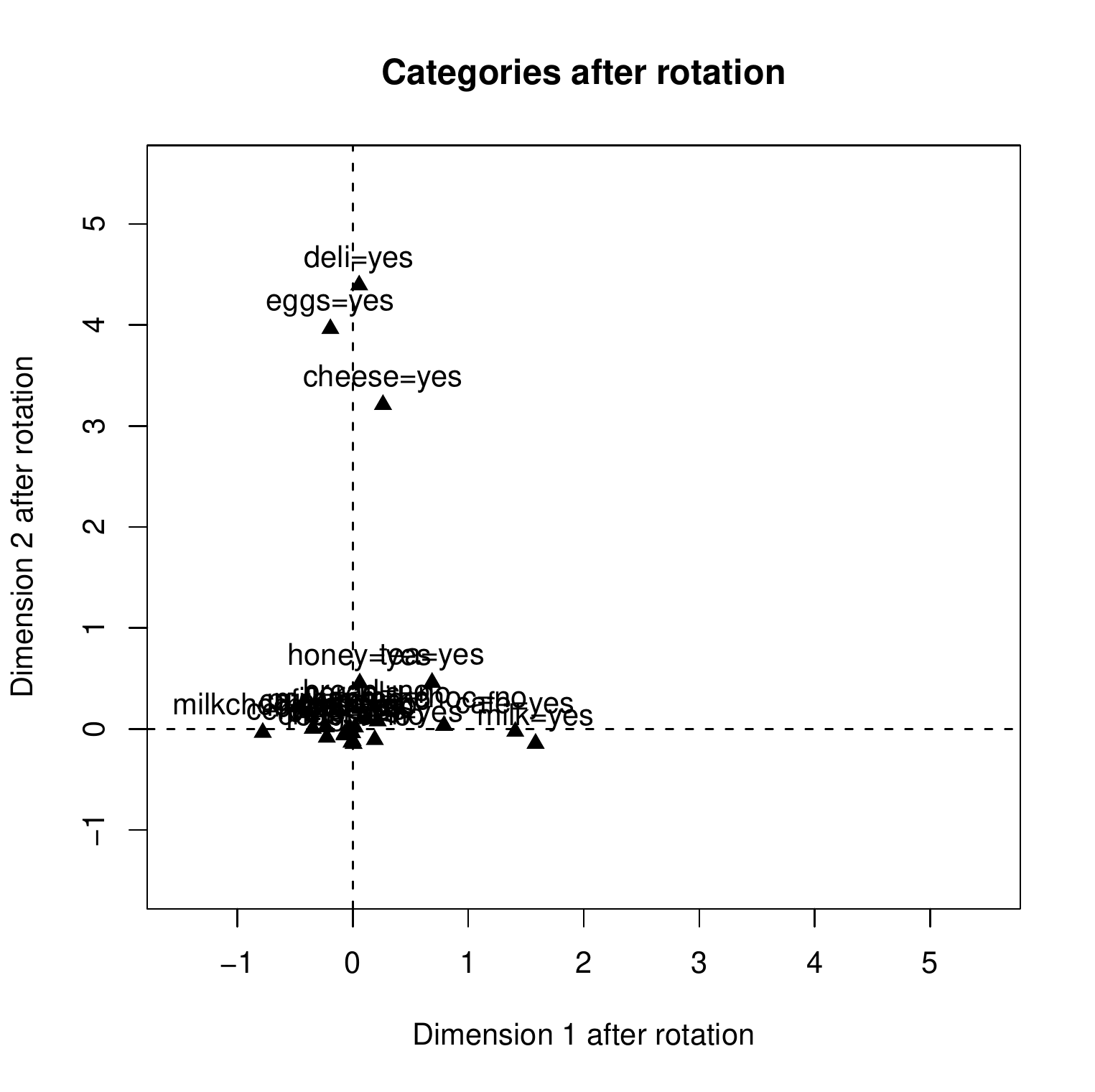}\\
\caption{Plots of the category coordinates on the first two components before and after rotation.}
\label{fig:categ}
\end{figure}

Note  that for binary variables  MCA and PCA lead to equivalent object scores and squared loadings (correlations are equal to correlation ratio). Then considering the data as quantitative in PCAMIX (equivalent to PCA in that case) gives the same results except for the plots of the categories which are not defined in that case.

\section{Conclusion}
We have given  in this paper a SVD based formulation of the PCAMIX method. This new formulation leads to an efficient procedure for varimax rotation in PCAMIX where a direct solution for the optimal angle of rotation $\theta$ has been obtained.  The numerical results have shown on simulations that this procedure is computationally more efficient than the procedure based on Kiers' matrix reformulation.  The numerical results have also shown on a real data application the interest of this algorithm in the context of MCA with graphical representations of both  variables and  categories after rotation. The PCAMIX procedure as well as the  rotation procedure have been implemented in the R package ``PCAmixdata''. 

\newpage

\paragraph{Appendix}~

Define the complex numbers:
\begin{equation}
\nonumber\label{AP1}
\begin{array}{lclclcl }
     a_{s} &\egaldef& a_{s,1}+ia_{s,2} \ ,& &  \tilde{a}_{s} &\egaldef& e^{-i\theta}a_{s} 
=\tilde{a}_{s,1}+i\tilde{a}_{s,2}
\ ,\\
     t_{j}   &\egaldef& \sum_{s\in I_{j}}a_{s}^{2} = {u_j}+i{v_j} \ ,& &  \tilde{t}_{j} &\egaldef & 
\sum_{s\in I_{j}}\tilde{a}_{s}^{2} =  e^{-2i\theta}t_{j} = \tilde{u}_{j}+i\tilde{v}_{j} \ ,
\end{array}
\end{equation}
where $\tilde{a}_{s,1},\tilde{a}_{s,2}$ have been defined in (\ref{as1}), $u_j,v_j$ in (\ref{usvd}), and where  $\tilde{u}_{j},\tilde{v}_{j}$ are given by the same formula as $u_j,v_j$, but with a tilde over $a_{s,1},a_{s,2} $.

We introduce now a complex-valued varimax function
 $F(\theta)$ of the rotation angle 
$\theta$ by:
\begin{equation}
\nonumber \label{AP2}
F(\theta) \egaldef p\sum_{j=1}^{p} \tilde{t}_{j}^{\,2} - (\sum_{j=1}^{p} \tilde{t}_{j} )^{2} = e^{-4i\theta}F(0) \ ,
\end{equation}
where $F(0)$ is simply obtained by suppressing the tilde in $F(\theta)$.
Development of $F(\theta)$ gives~:
\begin{equation}
\label{AP3}
   F(\theta) =   \underbrace{p\sum_{j=1}^{p}(\tilde{u}_{j}^{\,2}-\tilde{v}_{j}^{\,2}) 
     -(\sum_{j=1}^{p}\tilde{u}_{j})^{2}+(\sum_{j=1}^{p}\tilde{v}_{j})^{2} }_{g(\theta)}
     +\underbrace{2i \big\{ p\sum_{j=1}^{p}\tilde{u}_{j}\tilde{v}_{j}
     -\sum_{j=1}^{p}\tilde{u}_{j}\sum_{j=1}^{p}\tilde{v}_{j} \}}_{i\,h(\theta)} 
\end{equation}

 Comparison with the formula (\ref{rhopsi}), (\ref{akaiser}), (\ref{usvd}) defining $b,a,\rho,\psi$ shows that :
\begin{equation}
\nonumber \label{AP5}
  F(0)=g(0)+ih(0)=b+ia =\rho\, e^{i\psi} \ .
 \end{equation}
Hence :
\begin{equation}
\nonumber \label{AP6}
  F(\theta) =\rho\, e^{i(\psi-4\theta)} = \rho\,\big\{\cos(4\theta-\psi ) -i\sin(4\theta-\psi )\big\} \ .
\end{equation} 
But derivation of the varimax function $f(\theta)$ defined in (\ref{ftheta}) gives, using the fact that $a_{s,1}^{\prime}(\theta)=a_{s,2}(\theta)$ and $a_{s,2}^{\prime}(\theta)=-a_{s,1}(\theta)$ :
\begin{eqnarray}
\nonumber\label{dftheta1}
pf^{\prime}(\theta) & = & 2\big\{ p\sum_{j=1}^{p}\tilde{u}_{j}\tilde{v}_{j} -\sum_{j=1}^{p}\tilde{u}_{j}\sum_{j=1}^{p}\tilde{v}_{j}\big\} \\
\label{dftheta3}
    & = & h(\theta) = -\rho\,\sin(4\theta-\psi) \ , \\
\label{dftheta2}
    &=&  a\cos 4\theta -b \sin 4\theta \ ,
\end{eqnarray}
and  (\ref{dftheta3}) proves (\ref{ftheta2}) by integration.


\paragraph{References}~

\smallskip \noindent{de Leeuw, J., and Pruzansky, S., (1978), A new computational method to fit the weighted Euclidean distance model,  \textit{Psychometrika}, \textbf{43}, 479-490.}

 \smallskip \noindent{Escofier, B.,  (1979), Traitement simultan\'e de variables qualitatives et quantitatives en analyse factorielle [Simultaneous treatment of qualitative and quantitative variables in factor analysis], \textit{Cahiers de l'Analyse des Donn\'ees}, \textbf{4}, 137-146.}

 \smallskip \noindent{Jennrich, R.I.,  (2001), A simple general procedure for orthogonal rotation, \textit{Psychometrika}, \textbf{66}(2), 289-306.}

\smallskip \noindent{Kaiser, H.F., (1958), The varimax criterion for analytic rotation in factor analysis, \textit{Psychometrika}, \textbf{23}(3), 187-200.}

\smallskip \noindent{Kiers, H.A.L., (1991), Simple structure in Component Analysis Techniques for mixtures of qualitative and quantitative variables, \textit{Psychometrika}, \textbf{56}, 197-212.}

\smallskip \noindent{Neudecker, H., (1981), On the matrix formulation of Kaiser's varimax criterion, \textit{Psychometrika}, \textbf{46}, 343-345.}

\smallskip \noindent Pag\`es, J., (2004), Analyse Factorielle de donn\'ees mixtes [Factor Analysis for Mixed Data], {\it Revue de Statistique Appliqu\'ee}, {\bf 52}(4), 93-11.

\smallskip \noindent{ten Berge, J.M.F., (1984), A joint treatment of varimax rotation and the problem of diagonalizing symmetric matrices simultaneously in the least-squares sense, \textit{Psychometrika}, \textbf{49}, 347-358.}

\smallskip \noindent{ten Berge, J.M.F., (1995), Suppressing permutations or rigid planar rotations: a remedy against nonoptimal varimax rotations, \textit{Psychometrika}, \textbf{46} 60, 437-446}.

\end{document}